\documentstyle[12pt]{article}

\newlength{\dinwidth}
\newlength{\dinmargin}
\begin{document}

\def\bold#1{\setbox0=\hbox{$#1$}%
     \kern-.025em\copy0\kern-\wd0
     \kern.05em\copy0\kern-\wd0
     \kern-.025em\raise.0433em\box0 }
\def\slash#1{\setbox0=\hbox{$#1$}#1\hskip-\wd0\dimen0=5pt\advance
       \dimen0 by-\ht0\advance\dimen0 by\dp0\lower0.5\dimen0\hbox
         to\wd0{\hss\sl/\/\hss}}
\def\lq{\left [}
\def\rq{\right ]}
\def\LL{{\cal L}}
\def\VV{{\cal V}}
\def\AA{{\cal A}}
\def\MM{{\cal M}}

\newcommand{\be}{\begin{equation}}
\newcommand{\ee}{\end{equation}}
\newcommand{\bea}{\begin{eqnarray}}
\newcommand{\eea}{\end{eqnarray}}
\newcommand{\nn}{\nonumber}
\newcommand{\dd}{\displaystyle}
\newcommand{\bra}[1]{\left\langle #1 \right|}
\newcommand{\ket}[1]{\left| #1 \right\rangle}
\thispagestyle{empty}
\vspace*{1cm}
\rightline{BARI-TH/96-233}
\rightline{September 1996}
\vspace*{2cm}
\begin{center}
  \begin{Large}
  \begin{bf} 
Heavy quark kinetic energy  in  $B$ mesons\\
 by a QCD relativistic potential model\\
  \end{bf}
  \end{Large}
  \vspace{8mm}
  \begin{large}
Fulvia De Fazio \footnote{e-mail address:
DEFAZIO@BARI.INFN.IT}  \\
  \end{large}
  \vspace{6mm}
 Istituto Nazionale di Fisica Nucleare, Sezione di Bari, Italy\\
  \vspace{2mm}
 Dipartimento di Fisica, Universit\'a 
di Bari, Italy \\

\end{center}
\begin{quotation}
\vspace*{1.5cm}
\begin{center}
  \begin{bf}
  ABSTRACT
  \end{bf}
\end{center}
The matrix element of the kinetic energy operator 
 between $B$ 
meson states is computed by means of a QCD relativistic 
potential model, with the result:
$\mu_\pi^2=0.66 \hskip 3 pt GeV^2$. A comparison with the outcome of 
other theoretical
approaches and a discussion of the phenomenological implications of this 
result are carried out.
\vspace*{0.5cm}

\noindent
\end{quotation}

\newpage
\baselineskip=18pt
\setcounter{page}{1}
\section{Introduction}
In the last two decades, the study of hadronic processes involving heavy quarks 
 has attracted continuous interest both from experimental and 
theoretical sides. The main theoretical achievements have been obtained in the 
framework of Heavy Quark Effective Theory (HQET) \cite{isgur}, 
which describes the dynamics 
of heavy hadrons, i.e. hadrons containing a heavy quark $Q$, when $m_Q \to 
\infty$. The theory is based upon an effective lagrangian written in terms of 
effective fields, which is a systematic expansion in the inverse 
powers of the heavy quark mass $m_Q$. In particular, it has been pointed out 
that the expansion in the inverse powers of $m_Q$ is nothing else but an 
application of Operator Product Expansion (OPE) in the sector of heavy hadrons 
\cite{misha}.
The leading order effective Lagrangian displays heavy quark spin and flavour 
symmetries which are not present in full QCD. These symmetries are no longer
conserved at the next-to-leading order, and the 
${\cal O}\Big({1 \over m_Q}\Big)$ lagrangian reads as follows:
\be
{\cal L}={\bar h}_v i \; v \cdot D h_v+{1 \over 2 m_Q}
{\bar h}_v [D^2-(v \cdot D)^2] h_v
+{g_s \over 2 m_Q} {\bar h}_v {\sigma_{\mu \nu} G^{\mu \nu} \over 2} h_v
+{\cal O} \Big({1 \over m_Q^2}\Big) \hskip 3 pt .
\label{lag}
\ee
\noindent In the $m_Q \to \infty$ limit the field $h_v$ is related to the 
heavy quark field $Q$ by:
$Q=e^{-i m_Q v \cdot x} h_v(x)$, where $v_\mu$ is the heavy quark four-velocity 
which, in the $m_Q \to \infty$ limit, coincides with the one of the hadron 
\cite{georgi}.
In the hadron rest frame, the first of the two next-to-leading 
order operators appearing in (\ref{lag}) is the heavy quark non relativistic 
kinetic energy due to its residual motion, while the second one is the Pauli  
chromomagnetic interaction operator; 
they correspond in the Wilson expansion to dimension 5 
operators. Their matrix elements can be parametrized as follows:
\be
\mu_\pi^2(H_Q)={< H_Q|{\bar h}_v (i  {\vec {\cal D}})^2 h_v| H_Q > \over 2 
M_{H_Q}}
\label{mu_pi} \ee
\be
\mu_G^2(H_Q)={< H_Q|{\bar h}_v {g\over 2}  \sigma_{\mu \nu} G^{\mu \nu} h_v| 
H_Q > \over 2 M_{H_Q}}
\label{mu_g} \ee
\noindent where $H_Q$  denotes generically a hadron containing the heavy quark 
$Q$ and $ {\cal D}_\mu= {\partial}_\mu -i g  A_\mu$ is the 
covariant derivative. The normalization: $<H_Q|H_Q>=2 M_{H_Q}$ is understood.
\par
Since $\mu_G^2$ represents the chromomagnetic interaction between the heavy 
quark spin ${\vec s}_Q$ and the light cloud total angular momentum 
${\vec s}_\ell$, it can be obtained from the measured hyperfine mass splitting, 
when available. Its general expression reads: $\mu_G^2(H_Q)=
-2[J(J+1)-{3 \over 2}]\lambda_2$, 
where $J$ is the total spin and $\lambda_2$ is independent of the 
heavy quark mass.
Therefore, in case of $B$ mesons, we have:
\be
\mu_G^2(M_b)={d_M \over 4} (M_{B^*}^2-M_B^2)  \hskip 3 pt. 
\label{split} \ee
\noindent where $M_b=B,B^*$ and $d_M=3$ in the pseudoscalar case, $d_M=-1$ in 
the vector case (hence, from experimental data \cite{pdg}: 
$\mu_G^2(B) \simeq 0.36 \; GeV^2$,
$\mu_G^2(B^*) \simeq -0.12 \; GeV^2$).
On the other hand, it is expected to be zero for all baryons whose 
light  cloud is  in a ${\vec s}_\ell=0$ state, such as $\Lambda_Q$,  
$\Xi_Q$, while it should not vanish in the case of $\Omega_Q$ for which 
${\vec s}_\ell=1$, though it is not experimentally known, yet. Moreover, the 
mass splitting has been measured in the case of $\Sigma_b$: 
$M_{\Sigma_b^*}-M_{\Sigma_b}=56 \pm 16 \; MeV$ \cite{delphi}.

\par
$\mu_\pi^2$ represents the average square momentum carried by the heavy quark 
inside the hadron, that is, modulo a factor $2 m_Q$, its non relativistic 
kinetic energy. \par
These quantities are interesting for several reasons. Heavy hadrons 
masses are expected to scale with $m_Q$ as: 
\be
M_{H_Q}=m_Q+{\bar \Lambda}+{\mu_\pi^2 -   \mu_G^2 \over 2 m_Q}+ ...
\label{mass}
\ee
\noindent 
$\bar \Lambda$ represents the difference between the mass of 
the hadron and that of the heavy quark in  the $m_Q \to \infty$ limit. In this 
limit, it can be related to the trace anomaly  of QCD \cite{bsuv}: \\ 
${\bar \Lambda}={1 \over 2 M_{H_Q}}<H_Q| {\beta(\alpha_s) \over 4 \alpha_s} 
G^{\mu \nu} G_{\mu \nu}|H_Q>$, where $\beta$ is the Gell-Mann-Low function.
Moreover, if the inclusive semileptonic width of a heavy hadron is calculated 
by an expansion in the powers of ${1 \over m_Q}$, 
the following results are found: 
the leading term of the expansion coincides with the free quark decay rate 
(spectator model); no corrections of order ${1 \over m_Q}$ affect the rate; the 
${1 \over m_Q^2}$ corrections depend on $\mu_\pi^2$ and $\mu_G^2$ \cite{bigi}.
The absence of ${1 \over m_Q}$ corrections is sometimes referred to as CGG/BUV 
theorem.
As a consequence, these paramenters enter in 
the ratio of hadron lifetimes and in the 
lepton spectrum in inclusive transitions, which in principle 
are quantities directely 
comparable with experimental data \footnote{A critical analysis of such a 
procedure can be found in \cite{altarelli}.}. \par
It is worth noticing that $\mu_\pi^2$ and $\mu_G^2$ are the matrix elements of 
operators which are not sensitive to the light quarks flavour, i.e. they 
are $SU(3)$ singlet operators. The $SU(3)$ breaking effects emerge at 
${1 \over m_Q^3}$ 
level, due to four-quark operators; their matrix elements can be estimated by 
factorization in the case of mesons, and, in the case of baryons, by 
constituent quark models \cite{guberina} or field theoretical approaches, for 
example QCD sum rules \cite{noi}. \par
In this work we calculate $\mu_\pi^2$ in the case of $B$ mesons by means of a 
relativistic potential model. In the next section, 
after describing the relevant features of the model, we will present our 
results. Phenomenological implications and comparison with other approaches 
will be provided in section 3. Finally, we will draw our conclusions.

\section{Method and Results}
We will describe in the following the relativistic quark model used to 
compute $\mu_\pi^2$. Within this 
model, the state of a pseudoscalar
$(b {\bar q}_a)$ meson is written in terms of a wave function $\psi_B$
and of quark and antiquark creation operators; in the meson rest frame it 
reads:  
\be |B_a>=i {\delta_{\alpha \beta} \over \sqrt{3}} 
{\delta_{rs} \over \sqrt{2}} 
\int d \vec{k} \; \psi_B (\vec{k})\;
b^{\dag}(\vec{k}, r, \alpha) \; 
d^{\dag}_a(-\vec{k}, s, \beta)|0> \hskip 5 pt . \label{b} 
\ee
\noindent In (\ref{b}) $\alpha, \beta$ are colour indices, $r,s$ are spin 
indices and $a$ is a light flavour index; 
the operator $b^{\dag}$ creates the $b$ quark  with momentum: 
${\vec k} $, while $d^{\dag}_a$ creates a ${\bar q}_a$ antiquark  
with momentum
$-\vec{k} $.
The wave function $\psi_B$ is obtained as a solution of a 
Salpeter equation \cite{salpeter} which takes into account relativistic effects 
in the quark kinematics:
\be
\Big\{ \sqrt{\vec{k}^2 + m_b^2} +
\sqrt{\vec{k}^2 + m_{q_a}^2}-M_B \Big\}
\psi_B( \vec{k}) 
+ \int d \vec {k^{\prime}} V(\vec{k}, \vec{k^{\prime}}) 
  \psi_B (\vec{k^{\prime}})=0 \;\;\;. \label{7} \ee
\noindent Eq.
(\ref{7}) stems from the quark-antiquark 
Bethe-Salpeter equation in the approximation of  istantaneous interaction.
The interquark potential $V$ is represented by the Richardson potential
 \cite{richardson}, which 
reads in the $r-$space:
\be
V(r)={8 \pi \over 33 -2 n_f} \Lambda \Big[ \Lambda r-{f(\Lambda r) \over 
\Lambda r} \Big] \;\;\; ; \label{pot} 
\ee
\noindent $\Lambda$ is a parameter (chosen at the value $\Lambda=397 \;
MeV$ \cite{pietroni}), 
$n_f$ is the number of active flavours, and the function $f(t)$ is given by:
\be
f(t)={4 \over \pi} \int_0^\infty dq {sin(qt) \over q} \Big[ {1 \over ln(1+q^2)} 
- {1 \over q^2} \Big] \; . \label{f}
\ee
\noindent 
The potential (\ref{pot}) is linear 
at large distances in order to assure QCD confinement; at short distances it 
behaves as $- {\alpha_s(r) \over r}$, 
with $\alpha_s(r)$ logarithmically decreasing with the 
distance $r$ to reproduce the asymptotic freedom property of QCD.
Spin interaction effects are neglected since 
in the case of heavy mesons the chromomagnetic coupling is of order  
$m_Q^{-1}$.  The masses of the constituent quarks are fixed in such a 
way that the meson spectrum of the charmonium, the bottomomium and of the
heavy-light systems is reproduced: the fitted values are: $m_b= 4.89 \; GeV$, 
$m_q=m_u=m_d=0.038 \; GeV$. The mass of the charm quark has also been fixed:
$m_c=1.452 \; GeV$.
Finally, the wave function
$\psi_B$ is covariantly normalized: 
\be {1 \over (2\pi)^3} \int d \vec{k} |\psi_B(\vec k)|^2=
2M_B \hskip 3 pt . \label{eq : 8} \ee
\noindent
In the $B$ meson rest frame it is useful to define also 
the reduced wave function $u_B(k)$ ($k=|{\vec k}|$):
\be {u}_B(k)={k \; \psi_B(k) \over \sqrt{2} \pi } \hskip 3 pt  
\label{eq : 18} \ee
\noindent which  is normalized as: $\int_0^\infty dk |u_B(k)|^2=2 M_B$.
The function ${u}_B(k)$
can be obtained by numerically solving eq. (\ref{7}) 
using the Multhopp method described in \cite{multhopp,pietroni}.
 The $B-$ meson wave function, together with 
the wave functions computed for the other mesonic states analyzed within  this 
framework,  is the main outcome of the model; by using it
a number of hadronic
quantities characterizing the $B$ system have been computed, such as
semileptonic form factors, leptonic decay constants and 
strong coupling constants 
\cite{pietroni, pietro}.

By writing the heavy field, in the expression of the kinetic energy operator,
in terms of creation operators,  and by exploiting usual anticommutation 
relation and the normalization condition for the wave function, 
we obtain a simple expression for the $b$-quark average momentum squared in the 
$B$ meson:
\be
 K(m_b)={\int_0^\infty dk \; k^2 \; |u_B(k)|^2 \over 
\int_0^\infty dk \; |u_B(k)|^2} \hskip 3 pt . \label{p2} \ee
\noindent 
The parameter $\mu_\pi^2(B_d)$ should coincide with $ K(m_b)$ in 
the limit $m_b \to \infty$. In order to perform such limit, we compute 
$ K(m_b)$ 
for several values of $m_b$, using the wave function obtained in corrispondence 
to  the appropriate value of $m_b$; the result is
plotted in Fig. 1. We then extrapolate 
the resulting curve to $m_b \to \infty$ according to the expression:
\be
 K(m_b)=\mu_\pi^2(B_d) 
\Big( 1 +{ a \over m_b} \Big) \hskip 3 pt . \label{retta}
\ee
\noindent The result is:
\be 
\mu_\pi^2(B_d)=0.66 \; GeV^2 \hskip 3 pt  \label{ris}
\ee
\be
 a=-1.54 \; GeV \hskip 3 pt . \label{a} 
\ee
\noindent It should be noticed that this extrapolation procedure cannot be 
avoided in our approach since the resolution of the Salpeter equation can only 
be performed by numerical methods.\\
 Let us observe that the result: $a=-1.54 \; GeV$
 suggests that for the charm the ${1 \over m_Q}$ corrections play 
still an important role. On the other hand, the result obtained using the value 
of $m_b$ fixed within our model, i.e. $m_b=4.89 \; GeV$, is: $ K=0.46 \; 
GeV^2$, which means that for the $b$ quark the finite mass result differs from 
the asymptotic one at the level of $30  \%$. Moreover, eq. (\ref{ris})  
agrees  quite well with the QCD sum rule result \cite{ball}.\\
\noindent This result can be 
translated in a determination of the parameter ${\bar \Lambda}$. As a matter of 
fact, exploiting 
eqs. (\ref{split}) and (\ref{mass}) the following relation can be derived:
\be 
{\bar M}_B={3 M_{B^*}+M_B \over 4}=m_b +{\bar \Lambda}+{\mu_\pi^2 \over 2 m_b}
\label{mbbar}
\ee
\noindent which, using the value in eq.(\ref{ris}), gives: 
\be
{\bar \Lambda}=0.35 \; GeV 
\; . \label{lambda}
\ee
The results (\ref{ris}) and (\ref{lambda}) 
are also consistent with (\ref{mass}) if one neglects the higher order 
corrections in $1/m_b$ and uses the experimental value of $\mu_G^2(B_d)$.\\
Let us briefly discuss the  uncertainties of the result in (\ref{ris}).
They  mainly depend on the computed wave function $u_B(k)$, and, therefore, 
they can be estimated by modifying 
the shape of the wave function. We perform this analysis by comparing
the outcome in (\ref{ris}) with the result of a
similar constituent quark model, such as the one in 
ref. \cite{kim}, where  
the value of $\mu_\pi^2(B_d)$ is obtained by two independent
methods.  The first one consists in using the Altarelli et al. (ACCMM)
model \cite{accmm}, where the heavy 
quark momentum distribution inside the $B$ meson 
is assumed to be  gaussian: 
$\phi(|k|,P_F)=
{4 \over \sqrt{\pi} P_F^3} exp \Big( -{ |k|^2 \over P_F^2} \Big)$,
$P_F$ being a parameter. 
In terms of 
$P_F$ the kinetic energy $\mu_\pi^2(B_d)$ reads:
$\mu_\pi^2(B_d)={3 \over 2} P_F^2$. 
The value of $P_F$ has been obtained in \cite{kim} by a 
comparison with recent CLEO data \cite{cleo1} on the inclusive $B \to X \ell 
\nu$ semileptonic rate: 
 $P_F=0.54 \pm^{0.16}_{0.15} \hskip 3 pt 
GeV$, which corresponds to: $\mu_\pi^2 \simeq 0.44 \hskip 3 pt GeV^2$. 
The second method consists in using a quark model,
which gives 
$P_F=0.5-0.6 \hskip 3 pt GeV$, i. e.
$\mu_\pi^2 = 0.375-0.54 \hskip 3 pt GeV^2$.
We may compare our result with the range of values quoted 
above: $\mu_\pi^2=0.375- 0.54 \; GeV$, observing  that the comparison must be 
performed with our finite mass result, since the wave function used in 
\cite{kim} has been obtained for real values of $m_b$. By this comparison,
we can conservatively conclude that our result is 
affected by an error of $20 \% $  related to the shape of the $B$ meson 
wave function.

\section{Phenomenological implications}
Various determinations of the value of $\mu_\pi^2(B_d)$ exist 
in the literature
\footnote{ In the baryon sector, a result for the $\Lambda_b$ has been 
obtained by QCD sum rules: 
$\mu_\pi^2(\Lambda_b) \simeq 0.6 \; GeV^2$ \cite{paver}
with an estimated uncertainty of $30 \%$.};
they are collected in Table I.

The analyses in refs. \cite{kim}, \cite{lukesav}-\cite{chernyak}  consist 
in various attempts to extract or to put 
constraints on $\mu_\pi^2(B_d)$ 
from experimental data. In refs. \cite{lukesav},
 \cite{ligetinir} 
experimental data on semileptonic $B$ and $D$ decays are compared to 
theoretical predictions to extract $\mu_\pi^2$ as a function of $\bar \Lambda$
; in particular, in ref. \cite{lukesav} 
the QCD sum rule result \cite{neubert}: $\bar 
\Lambda=570 \pm 70 \; MeV$ is used to constrain $\mu_\pi^2$ in the 
range: $0.1 < \mu_\pi^2 \le 1.5 
\; GeV^2$. A similar approach is employed in ref. \cite{kapligeti}, 
where it is stressed 
the possibility of obtaining $\bar \Lambda$ and 
 $\mu_\pi^2$ from the moments of 
the photon spectrum in the decay $B \to X_s \gamma$. \\
QCD sum rules have been applied in refs.  \cite{shuryak} and
 \cite{ball} 
to determine $\mu_\pi^2$. Moreover, a great deal of works have been devoted to 
further 
constrain theoretically $\mu_\pi^2$. A field theoretical approach has been 
applied in ref. \cite{bsuv} (confirmed by a quantum mechanical 
approach in \cite{bigij,vol}) to state the inequality: $\mu_\pi^2 > 
\mu_G^2$ \footnote{In \cite{wise} the bound has been criticized, stating 
that the role of radiative corrections prevents any possibility of constraing 
$\mu_\pi^2$.}. Moreover, in \cite{grozin} a theoretical argument 
has been given to confirm and strengthen the bound, giving $\mu_\pi^2 > 0.45 \; 
GeV^2$ in the case of $B$ mesons. This argument is based upon the possibility 
of extracting $\mu_\pi^2$ from the slope of the Isgur-Wise function  which is 
related, at the leading $1/m_Q$ order, to the differential decay rate: ${d 
\Gamma \over d q^2} (B \to D^* \ell \nu)$ which can be obtained from 
experimental data \cite{gr19}. \\
Finally, very recently $\mu_\pi^2$ has 
been computed on the lattice \cite{lattice}.\\ On one hand,
this variety of results  suggests that further theoretical analyses 
of $\mu_\pi^2$ are required, on the other it shows that the experimental 
determinations are  a hard task. 
The main difficulty  lies in the smallness of the parameter $\mu_\pi^2$ and in 
the fact that it appears in physically measurable quantities always in 
connection with quantities that are determined in more or less broad range of 
values, such as ${\bar \Lambda}$ and the quark 
masses $m_c$, $m_b$. \par
As an example, we may consider 
the role played by $\mu_\pi^2$ in the $B$ semileptonic branching ratio, a 
well known problem in $B$ physics, since theoretical 
estimates are still larger than experimental data. The most recent 
experimental measurement  has been performed by CLEO Collaboration \cite{cleo1} 
giving: ${\cal B}(B \to X \ell \nu)=10.49 \pm 0.17 \pm 0.43 \; \%$. \\
From the point of view of the ${1 \over m_Q}$ expansion, the general procedure 
to determine an inclusive quantity  consists in applying the OPE to the forward 
matrix element of a weak transition operator \cite{misha}.
The resulting expression for the lepton spectrum in $B$ semileptonic decay, 
derived in \cite{misha1}, reads:
\begin{eqnarray}
{d \Gamma \over d y}&&= \Gamma_0 \; \theta (1-y-\rho)2 y^2 \Big\{ (1-f)^2(1+2f)
(2-y)+(1-f)^3(1-y) \nonumber \\
&&+ (1-f) \Big[ (1-f) \Big(2+{5 \over 3} y-2f+{10 \over 3}f y \Big) 
-{f^2 \over \rho} [2 y +f(12-12y+5y^2)] \Big]{\mu_G^2 \over m_b^2}  
\label{spettro} \\
&&- \Big[{5 \over 3}(1-f)^2(1+2f)y + {f^3 \over \rho} (1-f)(10y-8y^2)+ 
{f^4 \over \rho^2}(3-4f)(2y^2-y^3) \Big] {\mu_\pi^2 \over m_b^2} \Big\}
\; , \nonumber \end{eqnarray}
\noindent where: 
\be
\Gamma_0={G_F^2 m_b^5 \over 192 \pi^3}|V_{qb}|^2 \; , \hskip 1 cm 
f={\rho \over 1 -y}\; , \hskip 1 cm
\rho={m_q^2 \over m_b^2}\; , \hskip 1 cm 
y={2 E_\ell \over m_b} \; , \label{par}
\ee
\noindent and $m_q$ is the mass of the final quark $q$ 
\footnote{The expression of 
${\cal O}(\alpha_s)$ corrections to eq. (\ref{spettro}) 
can be found in \cite{jez}.}.\par
Let us consider the case $B \to X_c \ell \nu$. Eq. (\ref{spettro}) depends on 
the value of the charm quark mass $m_c$, which is fixed 
in the potential model to the value $m_c=1.452 \; GeV$ by
fitting  the whole charmonium spectrum .
 Using $V_{cb}=0.04$ and our result (\ref{ris}) we 
obtain the lepton spectrum displayed in fig. 2. This curve  must be compared to
the experimental distibution in \cite{cleo1}.
Since on the theoretical side, heavy quark mass 
expansion is unreliable for large lepton energies $E_\ell \ge 2 \; 
GeV$  \cite{misha1}, and, in the experimental analyses, 
high energy leptons must be selected in order to subtract the 
background of secondaries, the comparison theory-experiment 
can be performed only in
 a selected window of lepton energies. For example, as it has been done
 in refs. \cite{wise}, \cite{chernyak}, one can use the ratios:
\be
R_1={\int_{E_\ell \ge 1.5 \; GeV} E_\ell {d \Gamma \over d E_\ell}d E_\ell 
\over 
\int_{E_\ell \ge 1.5 \; GeV}  {d \Gamma \over d E_\ell}d E_\ell} \; ,
\hskip 1 cm 
R_2={\int_{E_\ell \ge 1.7 \; GeV}  {d \Gamma \over d E_\ell}d E_\ell \over 
\int_{E_\ell \ge 1.5 \; GeV}  {d \Gamma \over d E_\ell}d E_\ell} \; \label{r12}
\ee
\noindent where the dependence on the overall factor 
$|V_{cb}|^2 m_b^5$ cancels.
Using the experimental results: $R_1=1.7830$ 
$R_2=0.6108$  for the ratios in (\ref{r12}) 
the values of $\mu_\pi^2$
displayed in Table I have been obtained \cite{wise}, \cite{chernyak}
\footnote{The formulae in \cite{wise} for $R_{1,2}$ include also ${\cal O}(
\alpha_s)$ corrections.}. 
Using the formulae for $R_{1,2}$ in 
\cite{wise} and  our results (\ref{ris}), (\ref{lambda}), 
we obtain: $R_1=1.733$ $R_2=0.559$. 
However, as already pointed out in \cite{chernyak},  the parameters 
${\bar \Lambda}$, $\mu_\pi^2$ enter in $R_{1,2}$ as power corrections and 
represent a small effect in (\ref{r12}), so that very 
small changes in the theoretical or in the experimental expressions for 
$R_{1,2}$ would shift the values of 
${\bar \Lambda}$, $\mu_\pi^2$ towards very different results. 
One of such changes could be related to a different estimate of the secondary 
electron background, or, from the theoretical side, to the next order 
perturbative corrections, whose size is difficult to assess.
Therefore, one cannot avoid to conclude that the accuracy of 
a single determination 
from experimental data is difficult to check.
A set of independent measurements, from different channels, 
should be used, and an accurate cross check of the errors should 
be performed to detect the value of the parameter $\mu_\pi^2$.

\vskip 1cm

\noindent {\bf Acknowledgments\\} 
\noindent 
I thank P. Colangelo and G. Nardulli for interesting discussions.

\newpage

\begin{table}
\vspace{2cm}
\begin{tabular}{lcccccccc}
\hline \hline
& \cite{fls} &\cite{wise}& \cite{chernyak}&\cite{kim}& \cite{shuryak} &
\cite{ball}& \cite{lattice} &This paper \\ \hline
$\mu_\pi^2(B_d)$ &  0.1 & $0.19 \pm 0.1$ & 0.135 & $ 0.44 \pm 0.25$   
&$0.23 \pm 0.11$ & $0.60 \pm 0.10 $&$ -0.09 \pm 0.14$ & 0.66\\ \hline \hline
\end{tabular}
\end{table}
\vspace{1cm}
\noindent
{\bf Table I.} \\ \noindent
Results for the $B_d$ meson matrix element of the $b$ quark kinetic 
energy operator.

\newpage
\begin{center}
  \begin{Large}
  \begin{bf}
 Figure Captions
  \end{bf}
  \end{Large}
\end{center}
\vspace{5mm}
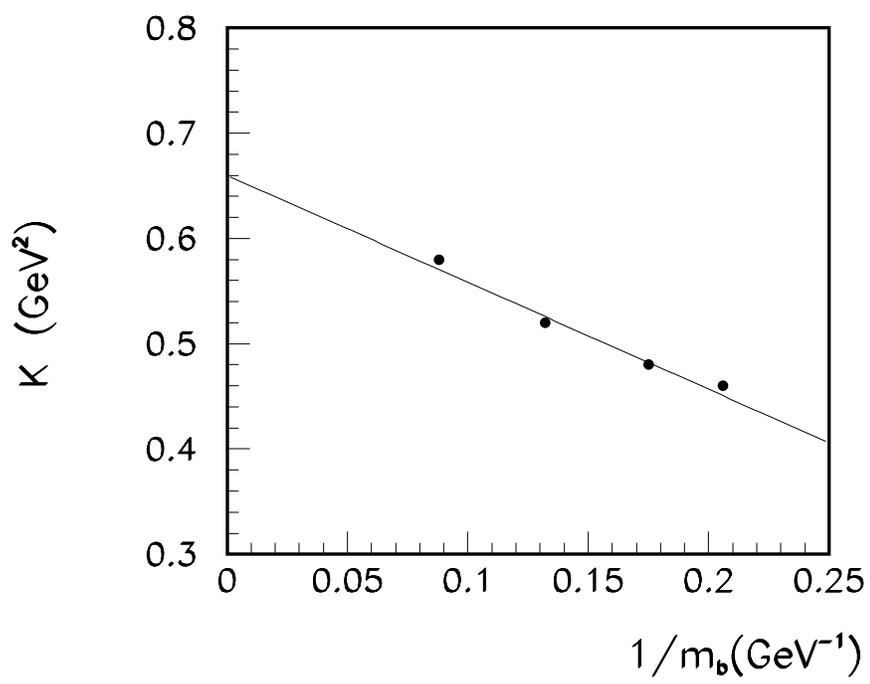
\noindent {\bf Figure 1}\\
\noindent
The $b$-quark average square momentum $K$ as a function of 
${1 \over m_b}$. The straight line results from a two parameters fit.
\par \noindent
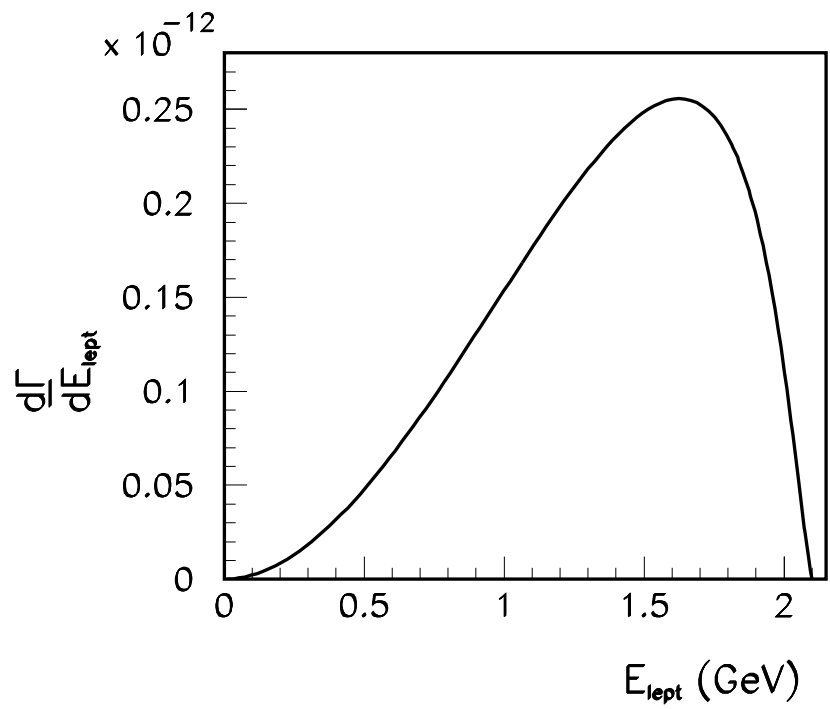
{\bf Figure 2}\\
\noindent
The lepton spectrum ${d \Gamma \over d E_\ell}$ in the semileptonic decay $B 
\to X_c \ell \nu$.
\end{document}